\documentclass{article}%
\usepackage{amsmath}
\usepackage{graphicx}%
\usepackage{amsfonts}%
\usepackage{amssymb}

\begin{document}

\title{Linear electric field frequency shift (important for next generation electric
dipole moment searches ) induced in confined gases by a magnetic field
gradient. }
\author{A.L. Barabanov$^{\#}$, R. Golub$^{+}$ and S.K. Lamoreaux$^{\ast}$\\$^{\#}$Kurchatov Institute\\\ 123182 Moscow, Russia\\$^{+}$Physics Department\\\ North Carolina State University\\\ Raleigh, NC 27606\\$^{\ast}$University of California,\\\ Los Alamos National Laboratory\\\ Physics Division\\\ Los Alamos NM 87545\\Yale University, Dept of Physics\\PO Box 208120, \\New Haven, CT 06520 }
\maketitle

\begin{abstract}
The search for particle electric dipole moments (edm) represents a most
promising way to search for physics beyond the standard model. A number of
groups are planning a new generation of experiments using stored gases of
various kinds. In order to achieve the target sensitivities it will be
necessary to deal with the systematic error resulting from the interaction of
the well-known $\overrightarrow{v}\times\overrightarrow{E}$ field with
magnetic field gradients (often referred to as the geometric phase effect
(Commins, ED; Am. J. Phys. \textbf{59}, 1077 (1991), Pendlebury, JM \emph{et
al;} Phys. Rev. \textbf{A70}, 032102 (2004)). This interaction produces a
frequency shift linear in the electric field, mimicking an edm. In this work
we introduce an analytic form for the velocity auto-correlation function which
determines the velocity-position correlation function which in turn determines
the behavior of the frequency shift (Lamoreaux, SK and Golub, R; Phys. Rev
\textbf{A71}, 032104 (2005)) and show how it depends on the operating
conditions of the experiment. We also discuss some additional issues.

\end{abstract}
\tableofcontents

\section{Introduction}

The proposition that the search for particle electric dipole moments (edm)
represents a reasonable method to \ look for physics beyond the standard model
\cite{weinb} is inspiring many groups to search for edm's in a variety of
systems. (See \cite{jmphinds} for a recent review). Experiments on several
systems including the neutron \cite{glrpp}, and several species of confined
gases \cite{gas} including Radium \cite{radium}, Radon \cite{radon} and Xenon
\cite{xenon} are in various stages of preparation. These experiments are all
hoping to reach sensitivities in the range of $10^{-27}-10^{-28}e-cm$.
\ Sensitivity in this range has already been achieved in the case of $Hg$
\cite{hg}. The experiments proposed represent a broad range of operating
conditions, from room temperature gases with buffer gas to laser cooled atoms
in a MOT.

In order to achieve the target sensitivities it will be necessary to deal with
the systematic error resulting from the interaction of the well-known
$\overrightarrow{v}\times\overrightarrow{E}$ field with magnetic field
gradients. Often referred to as the geometric phase effect \cite{cumm},
\cite{jmp} this interaction produces a frequency shift linear in the electric
field, mimicking an edm. This systematic effect is highly dependent on the
operating conditions of the experiment. While experiments in small vessels and
with high pressure buffer gas are expected to be relatively insensitive to the
systematic effect, each of the proposed experiments will have to be analyzed
in detail to judge its sensitivity to the effect and to find methods of
dealing with it.

In this work we introduce an analytic form of the correlation function which
determines the behavior of the frequency shift \cite{LG} and show in detail
how it depends on the operating conditions of the experiment. For clarity we
specialize the discussion to the Los Alamos proposal for a neutron edm search
using Ultra-cold neutrons (UCN) and $He^{3}$ atoms diffusing in superfluid
$He^{4}$ as a co-magnetometer, \cite{lanledm} but the generalization to other
cases is straightforward.

First analyzed by Commins \cite{cumm} in the context of a beam experiment, the
frequency shift has been discussed in some detail by Pendlebury et al
\cite{jmp} in connection with experiments involving stored particle gases.
Additional discussion and calculations have been given by \cite{LG}.

Our present understanding of the effect can best be summarized by figure 1,
which appeared as figure 3 in \cite{LG}. This is a plot of the normalized
(linear in $E$) frequency shift, $\delta\omega$, vs. normalized Larmor
frequency, $\omega_{0}$, for various values of collision mean free path,
$\lambda$, and wall specularity, calculated for particles moving with fixed
velocity, $v$, in a cylindrical measurement cell of radius $R.$ The horizontal
scale is fixed by the frequency of motion around the cell $\left(  v/R\right)
$. In general the shift for UCN will be given by a value of $\omega
_{o}/\left(  v/R\right)  >4$ while co-magnetometer atoms are characterized by
$\omega_{o}/\left(  v/R\right)  \ll1.$

These results have been obtained by numerical simulation of the
position-velocity correlation function and taking the Fourier transform.
According to [\cite{LG}, Eq. 26] the frequency shift is given by
\begin{equation}
\delta\omega=\frac{ab}{2}\lim_{\tau\rightarrow\infty}\int\limits_{0}^{\tau
}R\left(  t\right)  \cos(\omega_{0}t)dt, \label{2}%
\end{equation}
where $a=\frac{\gamma}{2}\partial B_{z}/\partial z,$ $b=$ $\gamma E/c,$
$\gamma$\emph{ }is the gyromagnetic ratio, and $R\left(  \tau\right)  $ is the
position-velocity correlation function for motion in the plane perpendicular
to the $z$ axis, defined in [\cite{LG}, Eq.. 27]:%
\begin{equation}
R\left(  \tau\right)  =\left\langle \overrightarrow{r_{\perp}}(t)\cdot
\overrightarrow{v_{\perp}}(t-\tau)-\overrightarrow{r_{\perp}}(t-\tau
)\cdot\overrightarrow{v_{\perp}}(t)\right\rangle . \label{1a}%
\end{equation}
From the experimental point of view it is very appealing to try to make use of
the zero crossing, apparent in figure 1, to reduce the effect.

In this note we present an analytic form for the velocity correlation function
from which $R\left(  \tau\right)  $ can be determined and compare it to the
results obtained previously by numerical simulations. In the collision-free
case we obtain the result obtained in (\cite{jmp}, Eq.. 78) for a single
trajectory. Using the analytic function for the case with gas collisions, we
average over a Maxwell velocity distribution and calculate the temperature
dependence of the frequency shift for $^{3}He$ diffusing in superfluid
$^{4}He$. We also propose a method for measuring the spectrum of the
correlation function, \emph{i.e.} the frequency dependence of the shift, directly.

\section{Analytical form for the correlation function $R\left(  \tau\right)  $}

According to \cite{LG} the correlation function $R(\tau)$ is determined by the
velocity autocorrelation function,
\begin{equation}
\psi(t)\equiv\langle\vec{v}_{\perp}(t)\cdot\vec{v}_{\perp}(0)\rangle,
\label{e0.1}%
\end{equation}
namely,
\begin{equation}
R(\tau)=2\int\limits_{0}^{\tau}\psi(t)dt. \label{e0.2}%
\end{equation}
Notice, that
\begin{equation}
R(\tau)\rightarrow0,\qquad\mbox{when}\qquad\tau\rightarrow\infty. \label{e0.3}%
\end{equation}
Thus we start by considering $\psi(t).$

\subsection{Specular wall collisions, no gas collisions}

\subsubsection{Specification of trajectories}

We consider particles moving ballistically in a cylindrical storage cell with
a fixed velocity $v$. As shown in \cite{jmp},\cite{LG} the frequency shift
depends only on the motion in the $x,y$ plane. Referring to figure 2, the
trajectory sweeps out an angle
\begin{equation}
\alpha=\arccos\left(  \frac{r}{R}\right)  \label{5b}%
\end{equation}
with respect to the center in a time
\begin{equation}
\tau_{w}=\frac{2R\sin\alpha}{v}\label{5c}%
\end{equation}
where $\tau_{w}$ is the time between wall collisions. For specular reflections
the angle $\alpha,$ characterizing the trajectory and the velocity are
unchanged by reflection.

The average angular velocity for a single trajectory is then%
\begin{equation}
\omega(\alpha)=\frac{2\alpha}{\tau_{w}}. \label{5d}%
\end{equation}

Let $F(r)\,dr$ be the probability that the trajectory of a particle has a
distance of closest approach to the center in the interval $(r,r+dr)$.
According figure 2,%
\begin{equation}
F(r)\,dr=\frac{2\sqrt{R^{2}-r^{2}}\,dr}{\pi R^{2}/2}\qquad\Rightarrow\qquad
F(r)=\frac{4\sqrt{R^{2}-r^{2}}}{\pi R^{2}}. \label{e4}%
\end{equation}
We will use the distribution $P(\alpha)d\alpha$ of the trajectories over the
angle $\alpha$, where
\begin{equation}
P(\alpha)|d\alpha|=F(r)|dr|\quad\Rightarrow\quad P(\alpha)=F(r)\left|
\frac{dr}{d\alpha}\right|  =\frac{4\sin^{2}\alpha}{\pi},\quad\int
\limits_{0}^{\pi/2}P(\alpha)d\alpha=1, \label{e5}%
\end{equation}
a result obtained in \cite{jmp}.

\subsubsection{The velocity autocorrelation function}

We now calculate the velocity autocorrelation function for the particles
moving along the trajectories with given $\alpha$ (or the pericenter
$r=R\cos\alpha$),%
\begin{equation}
\psi(\alpha,t)=\langle\vec{v}_{\perp}(t)\cdot\vec{v}_{\perp}(0)\rangle
,\qquad\psi(\alpha,0)=v_{\perp}^{2}. \label{e6}%
\end{equation}
Here the averaging goes only over the initial position of the particles. Let
\ the velocity autocorrelation function be denoted by
\begin{equation}
f(x,\alpha,t)\equiv\frac{\vec{v}_{\perp}(t)\cdot\vec{v}_{\perp}(0)}{v^{2}}.
\label{e7}%
\end{equation}
for a particle on a trajectory characterized by $\alpha$, starting at the
position $x,$ measured from the end of a chord, at $t=0.$ Thus,
\begin{equation}
\psi(\alpha,t)=\frac{v^{2}}{2R\sin\alpha}\int\limits_{0}^{2R\sin\alpha
}f(x,\alpha,t)dx. \label{e8}%
\end{equation}

As the speed $\left|  \overrightarrow{v}\right|  =v$ is not changed by the
collisions the velocity correlation function is proportional to $\cos
\theta(t)$ where $\theta(t)$ is the angle between $\vec{v}_{\perp}(t)$ and
$\vec{v}_{\perp}(0)$. The starting position $x$ determines the exact times at
which the collisions occur, the time between collisions being given by
(\ref{5c}) in all cases.

As a result, for the time $l\cdot\tau_{w}<t<(l+1)\cdot\tau_{w}$
($l=0,1,2\ldots$) the function $f$ is given by
\begin{equation}
f(x,\alpha,t)=\left\{
\begin{array}
[c]{ll}%
\cos(2(l+1)\alpha), & 0<x<v(t-l\cdot\tau_{w}),\\
\cos(2l\alpha), & v(t-l\cdot\tau_{w})<x<2R\sin\alpha,
\end{array}
\right.  \label{e9}%
\end{equation}
Then performing the averaging (\ref{e8}) the autocorrelation function takes
the form:%
\begin{equation}
\psi(\alpha,t)=v^{2}(A_{l}+B_{l}\frac{t}{\tau_{w}}), \label{e9b}%
\end{equation}
where%

\begin{equation}
A_{l}=(l+1)\cos(2l\alpha)-l\cos(2(l+1)\alpha), \label{e9c}%
\end{equation}%

\begin{equation}
B_{l}=\cos(2(l+1)\alpha)-\cos(2l\alpha). \label{e9d}%
\end{equation}

\subsubsection{Spectrum of the velocity correlation function}

The autocorrelation function can be written as a Fourier integral, valid for
both closed, periodic orbits and general open ones:%

\begin{equation}
\psi(\alpha,t)=\int\limits_{-\infty}^{+\infty}\Psi(\alpha,\omega)\cos(\omega
t)d\omega, \label{e10}%
\end{equation}
where $\Psi(\alpha,\omega)=\Psi^{\ast}(\alpha,\omega)$ and $\Psi(\alpha
,\omega)=\Psi(\alpha,-\omega)$, so that%
\begin{equation}
\Psi(\alpha,\omega)=\frac{1}{\pi}\int\limits_{0}^{\infty}\psi(\alpha
,t)\cos(\omega t)dt. \label{e11}%
\end{equation}

Straightforward calculation (see Appendix A) gives:%
\begin{align}
\Psi\left(  \alpha,\omega\right)   &  =\frac{2v^{2}\sin^{2}\alpha}{\omega
^{2}\tau_{w}^{2}}\sum_{m=-\infty}^{\infty}\left[  \delta\left(  \omega
+\frac{2\alpha-2\pi m}{\tau_{w}}\right)  +\delta\left(  \omega-\frac
{2\alpha+2\pi m}{\tau_{w}}\right)  \right] \nonumber\\
&  =\frac{2v^{2}\sin^{2}\alpha}{\omega^{2}\tau_{w}^{2}}\sum_{m=-\infty
}^{\infty}\left[  \delta\left(  \omega-\omega_{m}^{+}(\alpha)\right)
+\delta\left(  \omega-\omega_{m}^{-}(\alpha)\right)  \right]  \label{02a}%
\end{align}
where%

\begin{equation}
\omega_{m}^{\pm}(\alpha)=\frac{2\left(  \pi m\pm\alpha\right)  }{\tau_{w}}.
\label{e12b}%
\end{equation}
Notice that these frequencies are the resonant frequencies found previously in
the behavior of the the frequency shift as function of the angle $\alpha$ for
particles moving inside a cylindrical cell without any damping -- see figure 7
and equation (78) in \cite{jmp}. (Below we rederive this equation in the frame
of our approach and show how it can be generalized to account for damping).

\subsubsection{Solution for the frequency shift in the absence of gas collisions}

Now the frequency shift $\Delta\omega\left(  \alpha\right)  $ see equation
(40) \ ref. \cite{LG}
\begin{align}
-\Delta\omega\left(  \alpha\right)   &  =ab\int_{-\infty}^{\infty}\frac
{\Psi\left(  \alpha,\omega\right)  }{\left(  \omega_{o}^{2}-\omega^{2}\right)
}d\omega\nonumber\\
&  =2abv^{2}\sin^{2}\alpha\sum_{m=-\infty}^{\infty}\int_{-\infty}^{\infty
}\frac{\left[  \delta\left(  \omega+\frac{2\alpha-2\pi m}{\tau_{w}}\right)
+\delta\left(  \omega-\frac{2\alpha+2\pi m}{\tau_{w}}\right)  \right]
}{\left(  \tau_{w}\omega\right)  ^{2}\left(  \omega_{o}^{2}-\omega^{2}\right)
}d\omega\nonumber\\
&  =2abv^{2}\sin^{2}\alpha\sum_{m=-\infty}^{\infty}\tau_{w}^{2}\times
...\nonumber\\
&  ...\times\left[  \frac{1}{\left(  2\alpha-2\pi m\right)  ^{2}}\frac
{1}{\left(  \tau_{w}^{2}\omega_{o}^{2}-\left(  2\alpha-2\pi m\right)
^{2}\right)  }+\frac{1}{\left(  2\alpha+2\pi m\right)  ^{2}\left(  \tau
_{w}^{2}\omega_{o}^{2}-\left(  2\alpha+2\pi m\right)  ^{2}\right)  }\right]
\end{align}

The sum is over all $m$, so terms with $\pm m$ are included twice. Using
$\tau_{w}=2R\sin\alpha/v$ and writing $\omega_{o}\tau_{w}=2\omega_{o}%
R\sin\alpha/v=2\omega_{o}^{\prime}\sin\alpha\equiv2\delta_{o}$ with
$\omega_{o}^{\prime}=\omega_{o}R/v$ being the dimensionless frequency we then
find:
\begin{align}
-\Delta\omega\left(  \alpha\right)   &  =R^{2}ab\sin^{4}\alpha\sum_{m=-\infty
}^{\infty}\frac{1}{\left(  \alpha+\pi m\right)  ^{2}}\frac{1}{\left(
\delta_{o}^{2}-\left(  \alpha+\pi m\right)  ^{2}\right)  }\label{4a}\\
&  =\frac{1}{2\delta_{o}}R^{2}ab\sin^{4}\alpha\sum_{m=-\infty}^{\infty}%
\times...\nonumber\\
&  \left(  \frac{1}{\left(  \left(  \delta_{o}-\alpha\right)  -\pi m\right)
}+\frac{1}{\left(  \left(  \delta_{o}+\alpha\right)  +\pi m\right)  }\right)
\frac{1}{\left(  \alpha+\pi m\right)  ^{2}},\label{2a}\\
-\Delta\omega\left(  \alpha\right)   &  =R^{2}ab\sin^{2}\alpha\sum_{m=-\infty
}^{\infty}\frac{1}{\left(  \alpha+\pi m\right)  ^{2}}\frac{1}{\left(
\omega_{o}^{\prime2}-\frac{\left(  \alpha+\pi m\right)  ^{2}}{\sin^{2}\alpha
}\right)  }, \label{002}%
\end{align}
\ 

We now evaluate (\ref{2a}) using equation (\ref{b3}) derived in Appendix B.%
\[
\sum\limits_{n=-\infty}^{+\infty}\frac{1}{(\pi n-\varphi)^{2}(\pi n-\theta
)}=\frac{1}{(\varphi-\theta)\sin^{2}\varphi}-\frac{\cot\theta-\cot\varphi
}{(\varphi-\theta)^{2}}%
\]
Rewriting (\ref{2a}):%
\[
-\Delta\omega\left(  \alpha\right)  =\frac{1}{2\delta_{o}}R^{2}ab\sin
^{4}\alpha\Lambda
\]
with%
\begin{align*}
\Lambda &  =\sum_{m=-\infty}^{\infty}\left(  \frac{1}{\left(  \left(
\delta_{o}-\alpha\right)  -\pi m\right)  }+\frac{1}{\left(  \left(  \delta
_{o}+\alpha\right)  +\pi m\right)  }\right)  \frac{1}{\left(  \alpha+\pi
m\right)  ^{2}}\\
&  =\frac{1}{\delta_{o}\sin^{2}\alpha}+\frac{\cot\left(  \delta_{o}%
-\alpha\right)  +\cot\alpha}{\delta_{o}^{2}}+\frac{1}{\delta_{o}\sin^{2}%
\alpha}+\frac{\cot\left(  \delta_{o}+\alpha\right)  -\cot\alpha}{\delta
_{o}^{2}}\\
&  =\frac{2}{\delta_{o}\sin^{2}\alpha}+\frac{\cot\left(  \delta_{o}%
-\alpha\right)  +\cot\left(  \delta_{o}+\alpha\right)  }{\delta_{o}^{2}}\\
&  =\frac{2}{\delta_{o}\sin^{2}\alpha}\left(  1+\frac{\sin^{2}\alpha
\sin2\delta_{o}}{2\delta_{o}\sin\left(  \delta_{o}-\alpha\right)  \sin\left(
\delta_{o}+\alpha\right)  }\right)
\end{align*}
we find%
\begin{equation}
-\Delta\omega\left(  \alpha\right)  =\left(  \frac{v}{\omega_{o}}\right)
^{2}ab\left(  1+\frac{\sin^{2}\alpha\sin2\delta_{o}}{2\delta_{o}\sin\left(
\delta_{o}-\alpha\right)  \sin\left(  \delta_{o}+\alpha\right)  }\right)
\label{000}%
\end{equation}
This formula was originally derived in \cite{jmp}, (equ. 78) by direct
solution of the classical Bloch equations and the result shows the equivalence
of the two methods.

\subsection{Influence of Gas collisions}

\subsubsection{The collision-free velocity correlation function as a sum of
harmonic oscillators}

Substituting (\ref{02a}) into (\ref{e10}), we obtain for the velocity
autocorrelation function:%

\begin{align}
\psi(\alpha,t)  &  =\frac{2v^{2}\sin^{2}\alpha}{\tau_{w}^{2}}\sum_{n=-\infty
}^{\infty}\left[  \frac{\cos\omega_{n}^{+}\left(  \alpha\right)  t}{\left(
\omega_{n}^{+}\left(  \alpha\right)  \right)  ^{2}}+\frac{\cos\omega_{n}%
^{-}\left(  \alpha\right)  t}{\left(  \omega_{n}^{-}\left(  \alpha\right)
\right)  ^{2}}\right] \\
&  =\frac{2\sin^{2}\alpha}{\tau_{w}^{2}}\sum_{n=-\infty}^{\infty}\left[
\frac{\psi_{n}^{+}(\alpha,t)}{\left(  \omega_{n}^{+}\left(  \alpha\right)
\right)  ^{2}}+\frac{\psi_{n}^{-}(\alpha,t)}{\left(  \omega_{n}^{-}\left(
\alpha\right)  \right)  ^{2}}\right]  \label{e14}%
\end{align}
i.e. a sum of oscillating terms $\left(  \psi_{n}^{\pm}(\alpha,t)=v^{2}%
\cos\omega_{n}^{\pm}\left(  \alpha\right)  t\right)  $ with different
frequencies$.$ Each term obeys an equation:
\begin{equation}
\frac{d^{2}\psi_{n}^{\pm}(\alpha,t)}{dt^{2}}+(\omega_{n}^{\pm}(\alpha
))^{2}\psi_{n}^{\pm}(\alpha,t)=0. \label{17}%
\end{equation}

Notice that the fundamental frequency $\omega_{0}^{+}(\alpha)$ coincides with
(\ref{5d}). The corresponding term obviously dominates in the decomposition
(\ref{e14}).

\subsubsection{Velocity correlation function in the limit of short
gas-collision times}

We consider a particle that moves among scattering centers. If $\tau_{c}$ is
the average time between collisions the velocity autocorrelation function will
have the form \cite{mcgreg}, \cite{papoul}.
\begin{equation}
\psi(t)\equiv\langle\vec{v}(t)\vec{v}(0)\rangle=v^{2}\,e^{-\frac{t}{\tau_{c}}%
}. \label{4}%
\end{equation}
In the other words, $\psi(t)$ obeys the equation:
\begin{equation}
\frac{d\psi(t)}{dt}+\frac{1}{\tau_{c}}\psi(t)=0. \label{5a}%
\end{equation}

\subsubsection{Combined influence of gas and specular wall collisions}

The velocity autocorrelation function in the absence of collisions is given by
the sum of the motions of a group of harmonic oscillators (\ref{e14}). In the
presence of gas collisions the individual oscillators $\psi_{n}^{\pm}%
(\alpha,t)$ will obey the equation for a damped harmonic oscillator which is
the combination of (\ref{5a}) and (\ref{17}), i.e.%

\begin{equation}
\frac{d^{2}\psi_{n}(t)}{dt^{2}}+\frac{1}{\tau_{c}}\frac{d\psi_{n}(t)}%
{dt}+\omega_{n}^{2}\psi(t)=0, \label{e18}%
\end{equation}
with the initial condition:%

\begin{equation}
\psi_{n}^{\pm}(\alpha,0)=v^{2}. \label{e18b}%
\end{equation}
The boundary conditions (\ref{e0.2}) and (\ref{e0.3}) satisfied by $\psi(t)$
mean that%

\begin{equation}
\int\limits_{0}^{\tau}\psi_{n}(\alpha,t)dt\longrightarrow0,\qquad\mbox
{when}\qquad\tau\rightarrow\infty. \label{e19}%
\end{equation}
Generally, the equation for a damped harmonic oscillator (\ref{e18}) has the
general solution:
\begin{equation}
\psi(t)=c_{1}e^{-\eta_{1}t}+c_{2}e^{-\eta_{2}t}, \label{e25}%
\end{equation}
where
\begin{equation}
\eta_{1}=\frac{1}{2\tau_{c}}+\sqrt{\frac{1}{4\tau_{c}^{2}}-\omega^{2}%
}\,,\qquad\eta_{2}=\frac{1}{2\tau_{c}}-\sqrt{\frac{1}{4\tau_{c}^{2}}%
-\omega^{2}}\,. \label{e26}%
\end{equation}
Taking into account boundary conditions (\ref{e18b}) and (\ref{e19}), we get:
\begin{equation}
\psi_{n}(t)=\frac{v^{2}\eta_{1}}{\eta_{1}-\eta_{2}}\left(  e^{-\eta_{1}%
t}-\frac{\eta_{2}}{\eta_{1}}e^{-\eta_{2}t}\right)  . \label{e27}%
\end{equation}
The correlation function (\ref{e0.2}) takes the form (Eq. 36, \cite{LG}):%
\[
R_{n}(\alpha,\tau)=2\int\limits_{0}^{\tau}\psi_{n}(\alpha,t)dt=\frac{2v^{2}%
}{\eta_{1}-\eta_{2}}\left(  1-e^{-(\eta_{1}-\eta_{2})\tau}\right)
e^{-\eta_{2}\tau},
\]
The last expression can be rewritten in the form:
\begin{equation}
R_{n}(\alpha,\tau)=\frac{2\lambda v}{s(\omega,\tau_{c})}e^{-t/2\tau_{c}%
}\left(  e^{s(\omega,\tau_{c})t/2\tau_{c}}-e^{-s(\omega,\tau_{c})t/2\tau_{c}%
}\right)  , \label{e30}%
\end{equation}
with
\begin{equation}
\lambda=v\tau_{c},\qquad s(\omega,\tau_{c})=\sqrt{1-4\omega^{2}\tau_{c}^{2}}.
\label{e33}%
\end{equation}
We then have for a single oscillator (see equation \ref{2})%
\begin{equation}
\delta\omega_{n}=\frac{ab}{2}\lim_{t\rightarrow\infty}\int\limits_{0}^{t}%
R_{n}\left(  \tau\right)  \cos(\omega_{0}\tau)d\tau=abS_{n,d}\left(
\omega_{o}\right)
\end{equation}%

\begin{align}
S_{n,d}\left(  \omega_{o}\right)   &  =-v^{2}\frac{\left(  \omega_{o}%
^{2}-\omega_{n}^{2}\right)  }{\left(  \left(  \omega_{o}^{2}-\omega_{n}%
^{2}\right)  ^{2}+\frac{\omega_{o}^{2}}{\tau_{c}^{2}}\right)  }\nonumber\\
&  =-R^{2}\frac{\left(  \omega_{o}^{\prime2}-\omega_{n}^{\prime2}\right)
}{\left(  \left(  \omega_{o}^{\prime2}-\omega_{n}^{\prime2}\right)
^{2}+\omega_{o}^{2}r_{o}^{2}\right)  }%
\end{align}
where we again introduced $\omega_{o}^{\prime}=\omega_{o}R/v$ and
$r_{o}=R/\lambda_{c}(v)$ with $\lambda_{c}(v)$ the velocity dependent mean
free path. The frequency shift will then be a sum of such terms for each of
the oscillators as in (\ref{002}).

Comparing to the collision-free case in (\ref{002}) we see that in the
presence of damping the frequency shift will be given by
\begin{equation}
-\Delta\omega\left(  \alpha\right)  =R^{2}ab\sin^{2}\alpha\sum_{m=-\infty
}^{\infty}\frac{1}{\left(  \alpha+\pi m\right)  ^{2}}\left[  \frac{\left(
\omega_{o}^{\prime2}-\frac{\left(  \alpha+\pi m\right)  ^{2}}{\sin^{2}\alpha
}\right)  }{\left(  \left(  \omega_{o}^{\prime2}-\frac{\left(  \alpha+\pi
m\right)  ^{2}}{\sin^{2}\alpha}\right)  ^{2}+\omega_{o}^{\prime2}r_{o}%
^{2}\right)  }\right]  \label{003}%
\end{equation}
that is we go from the collision free case to the case of gas collisions by
replacing
\[
f_{\alpha}\left(  \omega^{\prime}\right)  =\frac{1}{\left(  \omega_{o}%
^{\prime2}-\frac{\left(  \alpha+\pi m\right)  ^{2}}{\sin^{2}\alpha}\right)  }%
\]
in (\ref{002}) by the square bracket in (\ref{003}) or by replacing
$f_{\alpha}\left(  \omega^{\prime}\right)  $ by $f_{\alpha}\left(
\omega^{\prime}\sqrt{1+i\frac{r_{o}}{\omega^{\prime}}}\right)  $ and taking
the real part. Since we have evaluated the summation (\ref{002}) we obtain the
frequency shift by making the equivalent transformation to (\ref{000})
\begin{equation}
-\Delta\omega\left(  \alpha\right)  =R^{2}ab\sin^{2}\alpha\operatorname{Re}%
\left\{  F_{P}(\alpha,\delta=\delta_{o}\sqrt{1+\frac{i}{\omega_{o}\tau_{c}}%
})\right\}  \label{004}%
\end{equation}
where
\[
F_{P}(\alpha,\delta)=\left(  1+\frac{\sin^{2}\alpha\sin2\delta}{2\delta
\sin\left(  \delta-\alpha\right)  \sin\left(  \delta+\alpha\right)  }\right)
\frac{1}{\delta^{2}}%
\]
(remember $\delta_{o}=\omega_{o}\tau_{w}/2$). For a fixed velocity we average
over $\alpha,$ according to (\ref{e5}):%
\begin{equation}
\Delta\omega=\int_{0}^{\pi/2}d\alpha P\left(  \alpha\right)  \Delta
\omega\left(  \alpha\right)  \label{005}%
\end{equation}
The results are shown in fig. 3 in comparison with the results of the
numerical simulations obtained in\cite{LG}.

We see the agreement is quite good, within the uncertainties of the numerical
simulations. The agreement in the region of the zero-crossings is excellent.

\section{Frequency shift averaged over velocity distribution, temperature dependence}

In the neutron edm experiment proposed by the EDM\ collaboration
\cite{lanledm} a dilute solution of $He^{3}$ dissolved in superfluid $He^{4}$
will be used as a co-magnetometer to monitor magnetic field fluctuations. As
such the $He^{3}$ will see essentially the same magnetic and electric fields
as the neutrons and will be subject to the linear E field systematic under discussion.

Using the analytical form of the correlation function we can average the $E$
field proportional frequency shift over the Maxwell velocity distribution for
a gas in thermal equilibrium. We take the realistic case of the mean free path
for collisions proportional to velocity, (collision time $\tau_{c}$,
independent of velocity). corresponding to a cross section $\sim1/v$. \ This
applies to $He^{3}$ in superfluid $He^{4}.$ Since the velocity of the $He^{3}$
is much less than the phonon velocity ($2.2\times10^{4}cm/\sec$) the collision
rate of phonons with the $He^{3}$ will be independent of the $He^{3}$
velocity. Thus in a time $\tau_{c}$ a $He^{3}$ with velocity $v$, will move a
distance $\lambda_{v}(T)=v\tau_{c}(T).$ We will obtain the collison time,
$\tau_{c}(T)$, from the measured values of the diffusion constant for $He^{3}$
in superfluid $He^{4}$ \cite{lametal}:%
\[
D\left(  T\right)  =\frac{1.6}{T^{7}}\ cm^{2}/\sec
\]

Then $\tau_{c}(T)=3D\left(  T\right)  /\left\langle v^{2}\right\rangle _{T}$
where $\left\langle v^{2}\right\rangle _{T}$ is the mean square velocity in a
volume of gas. $r_{o}$ is defined as
\[
r_{o}=\frac{R}{\lambda_{v}(T)}=\frac{R}{v\tau_{c}(T)}=\frac{R}{y\beta\left(
T\right)  \tau_{c}(T)}\equiv\frac{\overline{r}_{o}}{y}%
\]
with $y=v/\beta(T)$ and $\beta(T)$ is the most probable velocity in a volume.
$R$ the radius of the cylindircal vessel is taken as $R=25cm$ in the numerical
calculations. Both $\left\langle v^{2}\right\rangle _{T}$ and $\beta(T)$ are
calculated using the effective mass of $He^{3}$ in the superfluid: $m_{3}=7.2$ amu.

For a single velocity the frequency shift will be given by (\ref{003}) or
(\ref{004}) averaged over $\alpha$. Using Eq. (\ref{e5})%
\begin{align}
-\Delta\omega &  =\int_{0}^{\pi/2}d\alpha P\left(  \alpha\right)  \Delta
\omega\left(  \alpha\right) \nonumber\\
&  =R^{2}ab\frac{4}{\pi}\int_{0}^{\pi/2}d\alpha\sin^{4}\alpha\sum_{m}\frac
{1}{\left(  \alpha+\pi m\right)  ^{2}}\left[  \frac{\left(  \omega_{o}%
^{\prime2}-\frac{\left(  \alpha+\pi m\right)  ^{2}}{\sin^{2}\alpha}\right)
}{\left(  \left(  \omega_{o}^{\prime2}-\frac{\left(  \alpha+\pi m\right)
^{2}}{\sin^{2}\alpha}\right)  ^{2}+\omega_{o}^{\prime2}r_{o}^{2}\right)
}\right]
\end{align}
We now replace $\omega_{o}^{\prime}=\omega_{o}R/v=\omega_{o}R/y\beta
(T)=\omega_{o}^{\ast}/y$ where $\omega_{o}^{\ast}=\omega_{o}R/\beta(T)$. Then
\[
-\Delta\omega(y)=R^{2}ab\frac{4}{\pi}\int_{0}^{\pi/2}d\alpha\sin^{4}\alpha
\sum_{m}\frac{1}{\left(  \alpha+\pi m\right)  ^{2}}\left[  \frac{\left(
\omega_{o}^{\ast2}-\frac{\left(  \alpha+\pi m\right)  ^{2}}{\sin^{2}\alpha
}y^{2}\right)  y^{2}}{\left(  \omega_{o}^{\ast2}-\frac{\left(  \alpha+\pi
m\right)  ^{2}}{\sin^{2}\alpha}y^{2}\right)  ^{2}+\omega_{o}^{\ast2}%
\overline{r}_{o}^{2}}\right]
\]
where $\overline{r}_{o}=r_{o}y=R/\overline{\lambda}_{c}.$ Averaging over the
two dimensional velocity distribution ($\overrightarrow{v}_{\perp}$) we obtain
for the normalized frequency shift
\begin{equation}
\Psi\left(  \omega_{o}^{\ast},T\right)  =\frac{2}{abR^{2}}\int ye^{-y^{2}%
}\Delta\omega(y)dy
\end{equation}
The results are plotted in fig. 4 which shows the frequency shift as a
function of reduced frequency $\omega^{\ast}$ for various temperatures.

In figure 5 we show an expanded plot of the normalized frequency shift in the
region near the zero crossings as a function of temperature for fixed
$\omega^{\ast}\left(  T\right)  $.

It is evident that the collisional damping can lead to large reductions in the
effect for $He^{3}.$

\section{Non-specular wall collisions}

In this section we give a brief description of how non-specular wall
collisions can be included in the calculation. Detailed study of this problem
will be left for a future work.

In the preceding sections we have shown that the velocity autocorrelation
function can be regarded as the result of the sum of harmonic oscillators of
different frequencies.

Considering one such oscillator during one traversal of the cell the
oscillator will undergo a phase change
\begin{equation}
\phi=\omega\tau_{w}=2\alpha. \label{e33b}%
\end{equation}
A non-specular reflection from the wall would result in a change in the
incident angle for the next collision, $\chi$, by a random amount $\Delta\chi$
and hence a change in the accumulated oscillator phase by%
\begin{equation}
\Delta\phi=2\Delta\chi, \label{e33c}%
\end{equation}
because $\chi=\frac{\pi}{2}-\alpha$.

Since the changes $\Delta\phi$ are random the phase $\phi$ will make a random
walk so that after a time $t$ we will have\emph{ }%
\begin{equation}
\left\langle \left(  \Delta\phi\right)  ^{2}\right\rangle _{t}=4\left\langle
\left(  \Delta\chi\right)  ^{2}\right\rangle \frac{t}{\tau_{w}}. \label{e33d}%
\end{equation}
in the case of small $\Delta\phi\ll1$. Averaging the amplitude of the
oscillator over the distribution of $\Delta\phi,$ assuming a Gaussian
distribution for $\Delta\phi,$ the amplitude will be reduced by\emph{ }%
\begin{equation}
\left\langle \cos\phi\right\rangle \simeq e^{\left(  -\frac{1}{2}\left\langle
\left(  \Delta\phi\right)  ^{2}\right\rangle _{t}\right)  }\simeq\exp\left(
-\frac{t}{2\tau_{ns}}\right)  , \label{e33e}%
\end{equation}
where%

\begin{equation}
\frac{1}{\tau_{ns}}=\frac{4\left\langle \left(  \Delta\chi\right)
^{2}\right\rangle }{\tau_{w}}. \label{e33f}%
\end{equation}
\qquad\qquad\ 

Thus non-specular collisions can be taken into account by the change of the
damping term in (\ref{e18}):%

\begin{equation}
\frac{1}{\tau_{c}}\longrightarrow\frac{1}{\tau_{c}}+\frac{1}{\tau_{ns}}.
\label{e33g}%
\end{equation}
Non-specular wall collisions will thus have a different influence than the gas
collisions because of the dependence of $\tau_{w}\sim\sin\alpha$ on $\alpha$.

\section{Arbitrary Magnetic field geometry}

Our discussion has assumed a magnetic field configuration with $G_{z}=\partial
B_{z}/\partial z$ constant. Pendlebury et al \cite{jmp} have shown, using a
geometric phase argument, that regardless of the field geometry the effect
only depends on the volume average of $G_{z}$ in the high frequency (called by
them the adiabatic) limit. In a recent note, Harris and Pendlebury
\cite{harrjmp} have shown that in the case of a field produced by a dipole
external to the measurement cell, this does not hold in the low frequency
(diffusion) limit. In this section we discuss this problem using our
correlation function approach in order to give some physical insight into what
is happening and display details of the transition from one case to another.

\subsection{Short time (high frequency, adiabatic) limit of the correlation function}

Reference \cite{LG}, has shown that the systematic edm is given, in general,
as the Fourier transform of a certain correlation function of the time varying
field seen by the neutrons as they move through the apparatus. Equation (23)
of that paper gives the frequency shift proportional to $E$ as $\left(
\overrightarrow{\omega}\left(  t\right)  \text{ lies in the }x,y\text{
plane}\right)  $%

\begin{equation}
\delta\omega_{E}(t)=-\frac{1}{2}\int_{0}^{t}d\tau\left\{
\begin{array}
[c]{c}%
\cos\omega_{o}\tau\left[  \overrightarrow{\omega}\left(  t\right)
\times\overrightarrow{\omega}\left(  t-\tau\right)  \right] \\
+\sin\omega_{o}\tau\left[  \omega_{x}\left(  t\right)  \omega_{x}\left(
t-\tau\right)  +\omega_{y}\left(  t-\tau\right)  \omega_{y}\left(  t\right)
\right]
\end{array}
\right\}  \label{14}%
\end{equation}

It can be shown that the term multiplying $\sin\omega_{o}\tau$ goes to zero on
averaging over a uniform velocity distribution $\left(  \left\langle
v_{x}v_{y}\right\rangle =0,\quad v_{x}^{2}=v_{y}^{2}=v^{2}/2\right)  $ and
using $\overrightarrow{\nabla}\times\overrightarrow{B}=0.$Then, for short
times, $\tau,$%

\begin{align}
\delta\omega(t)  &  =-\frac{1}{2}\int_{0}^{t}d\tau\left\{  \cos\omega_{o}%
\tau\left[  \overrightarrow{\omega}\left(  t\right)  \times\left(
\overrightarrow{\omega}\left(  t\right)  -\frac{d\overrightarrow{\omega}}%
{dt}\tau+\frac{1}{2}\frac{d^{2}\overrightarrow{\omega}}{d\tau^{2}}\tau
^{2}+...\right)  \right]  \right\} \nonumber\\
&  =-\frac{1}{2}\int_{0}^{t}d\tau\left\{  \cos\omega_{o}\tau\left[
-\overrightarrow{\omega}\left(  t\right)  \times\left(  \frac{d\overrightarrow
{\omega}}{dt}\tau-\frac{1}{2}\frac{d^{2}\overrightarrow{\omega}}{d\tau^{2}%
}\tau^{2}+...\right)  \right]  \right\}  \label{15}%
\end{align}

We are considering values of $\tau$ so small that the velocity doesn't change
in that time interval $\left(  \tau<\tau_{coll}\right)  .$

Then
\begin{align*}
\overrightarrow{\omega}(t)  &  =\gamma\left(  \overrightarrow{B}%
_{xy}(t)+\overrightarrow{v}/c\times\overrightarrow{E}\right) \\
\frac{d\overrightarrow{\omega}}{dt}  &  =\gamma\left(  \overleftrightarrow
{\triangledown B}\left(  \overrightarrow{x}\left(  t\right)  \right)
\cdot\overrightarrow{v}\right) \\
\frac{d^{2}\overrightarrow{\omega}}{d\tau^{2}}  &  =\gamma\sum_{i,j}%
\frac{\partial^{2}\overrightarrow{B}}{\partial x_{i}\partial x_{j}}v_{i}v_{j}%
\end{align*}
and
\begin{equation}
\delta\omega(t)=-\frac{\gamma}{2c}\int_{0}^{t}d\tau\cos\omega_{o}\tau\left[
-\left(  \overrightarrow{B}_{xy}(t)+\overrightarrow{v}\times\overrightarrow
{E}\right)  \times\left(  \frac{\partial\overrightarrow{\omega}}{\partial
t}\tau-\frac{1}{2}\frac{\partial^{2}\overrightarrow{\omega}}{\partial\tau^{2}%
}\tau^{2}+...\right)  \right]  \label{12}%
\end{equation}
The term linear in $\overrightarrow{E\text{ }}$ and $\tau$ is then%
\begin{align}
\delta\omega(t)  &  =-\frac{\gamma^{2}}{2c}\int_{0}^{t}d\tau\cos\omega_{o}%
\tau\left[  \left(  \overleftrightarrow{\triangledown B}\cdot\overrightarrow
{v}\right)  \tau\times\left(  \overrightarrow{v}\times\overrightarrow
{E}\right)  \right] \nonumber\\
&  \equiv\frac{\gamma E}{2}\int_{0}^{t}d\tau\cos\omega_{o}\tau\left(
\alpha\tau\right)  \label{13}%
\end{align}
defining
\[
\alpha=\frac{\gamma}{c}\left(  \overleftrightarrow{\triangledown B}%
\cdot\overrightarrow{v}\right)  \cdot\overrightarrow{v}%
\]
We have now calculated the correlation function for short times. It starts at
zero at $\tau=0$ and rises as $\alpha\tau$. Eventually it will reach a
maximum. By concentrating on the high frequency $\left(  \omega_{o}\right)  $
behavior of $\delta\omega$ the result will be independent of the details of
the maximum, depending only on $\alpha.$ Thus we can replace $\alpha\tau$ in
(\ref{13}) by $\sin\alpha\tau$ or any function with the same initial slope.
Thus we are led to take
\begin{align*}
\delta\omega(t)  &  \equiv\frac{\gamma E}{2}\lim_{\omega_{o}\rightarrow\infty
}\int_{0}^{t}d\tau\cos\omega_{o}\tau\sin\alpha\tau\\
&  =\lim_{\omega_{o}\rightarrow\infty}\frac{\gamma E}{2}\frac{\alpha}%
{\omega_{o}^{2}-\alpha^{2}}=\frac{\gamma E}{2}\frac{\alpha}{\omega_{o}^{2}}\\
&  =\frac{E}{2cB_{o}^{2}}\left(  \overleftrightarrow{\triangledown B}%
\cdot\overrightarrow{v}\right)  \cdot\overrightarrow{v}%
\end{align*}
Introducing components, taking averages and using $\overrightarrow
{\triangledown}\cdot\overrightarrow{B}=0$ this reduces to%

\begin{equation}
\overline{\delta\omega}_{geo}=-Ev^{2}\frac{1}{4cB_{o}^{2}}\left\langle
\frac{\partial B_{z}}{\partial z}\right\rangle
\end{equation}
in agreement with Eq.. $\left(  2\right)  $ of \cite{LG} if, in that equation,
$R^{2}\omega_{r}^{2}$ is replaced by $\left\langle v^{2}\right\rangle
=v^{2}/2$. We have shown that in the adiabatic (short time) limit the
systematic (false) edm effect depends only on $\left\langle \frac{\partial
B_{z}}{\partial z}\right\rangle $ regardless of the geometry of the magnetic
field, a result obtained previously by Pendlebury et al \cite{jmp} and
confirmed in \cite{harrjmp}.

The next order term in (\ref{12}) is easily seen to be of order $v^{3}\tau
^{2}$ and so will average to zero, the next term which contributes will be of
order $v^{4}\tau^{3}$ and so will be negligible in the short time limit we are
considering. The condition for this to be valid is $\left(  v\tau/L\right)
^{2}\ll1$ where $L$ is the scale of variations in the applied magnetic field
$\left(  \frac{\partial B_{z}}{\partial z}\frac{1}{B_{z}}\sim L^{-1}\right)  $

\subsection{Longer time behavior of the correlation function}

For long times the expansion (\ref{15}) is clearly not valid and we must
expand in a series in the spatial coordinates. We start from%
\[
\delta\omega=-\frac{\gamma^{2}}{2}\int d\tau\cos\omega_{o}\tau\left\langle
\overrightarrow{B}^{\prime}\left(  t\right)  \times\overrightarrow{B}^{\prime
}\left(  t-\tau\right)  \right\rangle _{z}%
\]
where $b=E/c$, the brackets represent an ensemble average and
\begin{align*}
B_{x}^{\prime}  &  =B_{x}\left(  \overrightarrow{r}\left(  t\right)  \right)
-bv_{y}\\
B_{y}^{\prime}  &  =B_{y}\left(  \overrightarrow{r}\left(  t\right)  \right)
+bv_{x}%
\end{align*}
Then we write%
\begin{equation}
\left\langle \overrightarrow{B}^{\prime}\left(  t\right)  \times
\overrightarrow{B}^{\prime}\left(  t-\tau\right)  \right\rangle _{z}%
=b\left\langle
\begin{array}
[c]{c}%
B_{x}\left(  \overrightarrow{r}\left(  t\right)  \right)  v_{x}\left(
t-\tau\right)  -v_{y}\left(  t\right)  B_{y}\left(  \overrightarrow{r}\left(
t-\tau\right)  \right) \\
-\left(  B_{x}\left(  \overrightarrow{r}\left(  t-\tau\right)  \right)
v_{x}\left(  t\right)  -v_{y}\left(  t-\tau\right)  B_{y}\left(
\overrightarrow{r}\left(  t\right)  \right)  \right)
\end{array}
\right\rangle \label{11}%
\end{equation}
and expand the field in a Taylor series%

\begin{align*}
B_{x}\left(  \overrightarrow{r}\left(  t\right)  \right)  =  &  \left(
B_{x}\left(  0,0,0,t\right)  +\left.  \frac{\partial B_{x}}{\partial
x}\right|  _{o}x\left(  t\right)  +\left.  \frac{\partial B_{x}}{\partial
y}\right|  _{o}y\left(  t\right)  +\left.  \frac{\partial B_{x}}{\partial
z}\right|  _{o}z\left(  t\right)  \right)  +\\
&  +\left(  \left.  \frac{\partial^{2}B_{x}}{\partial x^{2}}\right|  _{o}%
x^{2}\left(  t\right)  +\left.  \frac{\partial^{2}B_{x}}{\partial y^{2}%
}\right|  _{o}y^{2}\left(  t\right)  +\left.  \frac{\partial^{2}B_{x}%
}{\partial z^{2}}\right|  _{o}z^{2}\left(  t\right)  \right) \\
&  +\left(  \left.  \frac{\partial^{2}B_{x}}{\partial x\partial y}\right|
_{o}y(t)x\left(  t\right)  +\left.  \frac{\partial^{2}B_{x}}{\partial
y\partial z}\right|  _{o}z\left(  t\right)  y\left(  t\right)  +\left.
\frac{\partial^{2}B_{x}}{\partial z\partial x}\right|  _{o}x\left(  t\right)
z\left(  t\right)  \right) \\
&  +\left(  \left.  \frac{\partial^{3}B_{x}}{\partial x^{3}}\right|  _{o}%
x^{3}\left(  t\right)  +\left.  \frac{\partial^{3}B_{x}}{\partial y^{3}%
}\right|  _{o}y^{3}\left(  t\right)  +....\right)
\end{align*}
(similarly for $B_{y}$). Concentrating on the first and last terms in
(\ref{11}) and noting that there are no correlations between any functions
$f(x_{i},v_{i})$ and $g\left(  x_{j},v_{j}\right)  $ we find%
\[
\sum_{x_{i}=x,y}\left\langle \left(  \left.  \frac{\partial B_{x_{i}}%
}{\partial x_{i}}\right|  _{o}x_{i}\left(  t\right)  +\left.  \frac
{\partial^{2}B_{x_{i}}}{\partial x_{i}^{2}}\right|  _{o}x_{i}^{2}\left(
t\right)  +\left.  \frac{\partial^{3}B_{x}}{\partial x_{i}^{3}}\right|
_{o}x_{i}^{3}\left(  t\right)  ...\right)  v_{x_{i}}\left(  t-\tau\right)
-\left\{  \left(  t\right)  \Leftrightarrow\left(  t-\tau\right)  \right\}
\right\rangle
\]
where the second term is obtained from the first by interchanging $\left(
t\right)  $ and $\left(  t-\tau\right)  $.

By symmetry we see that $\left\langle x_{i}^{2}\left(  t\right)  v_{x_{i}%
}\left(  t-\tau\right)  \right\rangle =0$ so that the next contributing term
will be proportional to
\[
\left.  \frac{\partial^{3}B_{x}}{\partial x_{i}^{3}}\right|  _{o}\left\langle
x_{i}^{3}\left(  t\right)  v_{x_{i}}\left(  t-\tau\right)  \right\rangle
\]
The first order term will be proportional to
\[
\frac{\partial B_{x}}{\partial x}+\frac{\partial B_{y}}{\partial y}%
=-\frac{\partial B_{z}}{\partial z}%
\]
We see that the condition
\[
\left.  \frac{\partial^{3}B_{x_{i}}}{\partial x_{i}^{3}}\right|  _{o}R^{2}%
\ll\left.  \frac{\partial B_{x_{i}}}{\partial x_{i}}\right|  _{o}%
\quad\emph{or\quad}\frac{R^{2}}{L^{2}}\ll1
\]
will insure that the higher order terms can be neglected. In the extreme case
considered by Harris and Pendlebury, \cite{harrjmp}, this condition is
strongly violated so our method cannot be applied since the higher order terms
remain significant.

\section{Discussion}

We have derived the analytic form of the (2 dimensional) velocity
autocorrelation function (vcf) for particles moving in a specularly reflecting
cylindrical vessel. As the vcf determines the position-velocity correlation
function (pvcf) we have calculated the Fourier transform of this latter
function thus obtaining the frequency shift for a single group of trajectories
characterized by the angle $\alpha$. Our results duplicate the analytic result
found in \cite{jmp} by direct solution of the classical Bloch equations. We
then show how the analytic formula for single trajectories can be extended to
take into account gas collisions and obtain\ (after integrating over
trajectories) the linear in $E$ frequency shift for a single velocity which
has been studied previously, our results agreeing with those obtained by
numerical simulation of the pvcf with collisions in \cite{LG}. We then perform
an average over the Maxwell velocity distribution using the temperature
dependent collision times appropriate for $He^{3}$ diffusing in superfluid
$He^{4}$ and obtain the temperature dependence of the frequency shift in this
case. Figures 4 and 5, which are based on an exact average over the Maxwell
distribution of the $He^{3}$ velocities, imply that one should be able to
control the effect to high degree.

Due to the heavy mass and slow velocity of the $He^{3}$, Baym and Ebner
\cite{baymebn} conclude that the phonon scattering on $He^{3}$ is
predominantly elastic. Single phonon absorption is kinematically forbidden on
a single $He^{3}$ and can only take place as a result of $He^{3}-He^{3}$
collisions which will be negligible for the low $He^{3}$ densities considered
here. Thus our approach, where we calculate the correlation function for an
ensemble of trajectories with constant $He^{3}$ velocity and then average the
result over the velocity distribution should be an excellent approximation.

In addition we have discussed the conditions under which the frequency shift
can be shown to depend only on the volume average of $\partial B_{z}/\partial
z$.

\section{Appendix A}

For a general orbit the velocity autocorrelation function is not periodic.
However, using the form (\ref{e9b}) for the correlation function, we can
calculate its Fourier transform straighforwardly:%

\begin{equation}
\Psi(\alpha,\omega)=\frac{v^{2}}{\pi}\sum\limits_{l=0,1,2\ldots}\left(
A_{l}\int\limits_{l\tau_{w}}^{(l+1)\tau_{w}}\cos(\omega t)dt+B_{l}%
\int\limits_{l\tau_{w}}^{(l+1)\tau_{w}}\frac{t}{\tau_{w}}\cos(\omega
t)dt\right)  . \label{a22}%
\end{equation}
Now ($\delta=\omega\tau_{w}$)%
\begin{equation}
\int\limits_{l\tau_{w}}^{(l+1)\tau_{w}}\!\!\!\!\cos(\omega t)dt=\frac
{\sin((l+1)\delta)-\sin(l\delta)}{\omega}, \label{aa}%
\end{equation}%

\begin{equation}
\int\limits_{l\tau_{w}}^{(l+1)\tau_{w}}\!\!\!\!\frac{t}{\tau_{w}}\cos(\omega
t)dt=\frac{(l+1)\sin((l+1)\delta)-l\sin(l\delta)}{\omega}+\frac{\cos
((l+1)\delta)-\cos(l\delta)}{\omega\delta}, \label{ab}%
\end{equation}
We separate out the terms in (\ref{a22}), according to whether they come from
sines or cosines in\ (\ref{aa}) or (\ref{ab}).

Taking the sine terms first we have
\begin{align}
&  \left(  \left(  l+1\right)  \cos(2l\alpha)-l\cos(2\left(  l+1\right)
\alpha)\right)  \left(  \sin(\left(  l+1\right)  \delta)-\sin(l\delta)\right)
-\nonumber\\
&  \left(  \cos(2\left(  l+1\right)  \alpha)-\cos2l\alpha\right)  \left(
\left(  l+1\right)  \left(  \sin(\left(  l+1\right)  \delta\right)
-l\sin(l\delta)\right) \nonumber\\
&  =\sin(\left(  l+1\right)  \delta)\left(  \cos2\left(  l+1\right)
\alpha\right)  -\sin(l\delta)\cos(2l\alpha),
\end{align}
the sine terms in $\Psi\left(  \alpha,\omega\right)  $ are
\begin{align}
\Psi_{s}\left(  \alpha,\omega\right)   &  =\frac{v^{2}}{\pi\omega}\sum
_{l=0}^{\infty}\left(  \sin(\left(  l+1\right)  \delta)\cos(2\left(
l+1\right)  \alpha)-\sin(l\delta)\cos(2l\alpha)\right) \nonumber\\
&  =\frac{v^{2}}{\pi\omega}\sum_{l=0}^{N}(f\left(  l+1\right)  -f(l))=\frac
{v^{2}}{\pi\omega}\left(
\begin{array}
[c]{c}%
f\left(  1\right)  -f\left(  0\right)  +\\
f\left(  2\right)  -f\left(  1\right)  +\\
\ldots+\\
f\left(  N\right)  -f\left(  N-1\right)  +\\
f\left(  N+1\right)  -f\left(  N\right)
\end{array}
\right) \nonumber\\
&  =\frac{v^{2}}{\pi\omega}f\left(  N+1\right)  =\frac{v^{2}}{\pi\omega}%
\sin(\left(  N+1\right)  \delta)\cos(2\left(  N+1\right)  \alpha).
\end{align}

Turning to the cosine terms we have%
\begin{align}
\Psi_{c}\left(  \alpha,\omega\right)   &  =\frac{v^{2}}{\pi\omega\delta}%
\sum_{l=0}^{\infty}B_{l}\left(  \cos(\left(  l+1\right)  \delta)-\cos
(l\delta)\right)  =\nonumber\\
&  =\frac{v^{2}}{\pi\omega\delta}\left(  \sum_{l=0}^{N}B_{l}\cos(\left(
l+1\right)  \delta)-\sum_{l=0}^{N}B_{l}\cos(l\delta)\right) \nonumber\\
&  =\frac{v^{2}}{\pi\omega\delta}\left(  B_{N}\cos(\left(  N+1\right)
\delta)-B_{0}+\sum_{l=1}^{N}(B_{l-1}-B_{l})\cos(l\delta)\right)  ,
\end{align}
where%

\begin{align*}
B_{l-1}-B_{l}  &  =2\cos(2l\alpha)-\cos(2(l-1)\alpha)-\cos(2(l+1)\alpha)\\
&  =2\cos(2l\alpha)(1-\cos(2\alpha))=4\sin^{2}\alpha\cos(2l\alpha).
\end{align*}
Then
\begin{equation}
\Psi_{c}\left(  \alpha,\omega\right)  =\frac{v^{2}}{\pi\omega\delta}\left(
1-\cos2\alpha+B_{N}\cos(\left(  N+1\right)  \delta)+4\sin^{2}\alpha\sum
_{l=1}^{N}\cos(l\delta)\cos(2l\alpha)\right)  ,
\end{equation}%
\begin{align}
h(\delta,\alpha)  &  =\sum_{l=1}^{N}\cos(l\delta)\cos(2l\alpha)=\frac{1}%
{2}\sum_{l=1}^{N}(\cos(l\left(  \delta+2\alpha\right)  )+\cos(l\left(
\delta-2\alpha\right)  ))\nonumber\\
&  =\frac{1}{2}\left(  \frac{\sin\left(  \left(  N+\frac{1}{2}\right)  \left(
\delta+2\alpha\right)  \right)  }{2\sin\frac{\left(  \delta+2\alpha\right)
}{2}}+\frac{\sin\left(  \left(  N+\frac{1}{2}\right)  \left(  \delta
-2\alpha\right)  \right)  }{2\sin\frac{\left(  \delta-2\alpha\right)  }{2}%
}-1\right)
\end{align}
except when $\delta\pm2\alpha=2\pi n$ in which case $h(\delta,\alpha)=N/2.$

So (we will take the limit as $N\rightarrow\infty$)%

\[
\lim_{N\rightarrow\infty}\frac{\sin(N+\frac{1}{2})x}{\sin\frac{1}{2}x}%
=\lim_{N\rightarrow\infty}\left(  \frac{\sin Nx\cos\frac{1}{2}x}{\sin\frac
{1}{2}x}+\cos Nx\right)  =\lim_{N\rightarrow\infty}\left(  \frac{\sin
Nx\cos\frac{1}{2}x}{\sin\frac{1}{2}x}\right)  =0
\]
except when $x=2\pi m$,%

\begin{equation}
\lim_{N\rightarrow\infty}\left(  \frac{\sin Nx\cos\frac{1}{2}x}{\sin\frac
{1}{2}x}\right)  =\sum_{m=-\infty}^{\infty}2\pi\delta\left(  x-2\pi m\right)
.
\end{equation}
As a result (\ref{a22}) becomes ($\delta=\omega\tau_{w}$)%

\begin{equation}
\Psi\left(  \alpha,\omega\right)  =\frac{v^{2}}{\pi\omega^{2}\tau_{w}}\sin
^{2}\alpha\sum_{m=-\infty}^{\infty}\left(  2\pi\delta\left(  \omega\tau
_{w}+2\alpha-2\pi m\right)  +2\pi\delta\left(  \omega\tau_{w}-2\alpha-2\pi
m\right)  \right)  .
\end{equation}
Using $\delta\left(  \omega\tau_{w}+2\alpha-2\pi m\right)  =\frac{1}{\tau_{w}%
}\delta\left(  \omega+\frac{2\alpha-2\pi m}{\tau_{w}}\right)  $ we have
finally
\begin{align}
\Psi\left(  \alpha,\omega\right)   &  =\frac{2v^{2}\sin^{2}\alpha}{\omega
^{2}\tau_{w}^{2}}\sum_{m=-\infty}^{\infty}\left(  \delta\left(  \omega
+\frac{2\alpha-2\pi m}{\tau_{w}}\right)  +\delta\left(  \omega-\frac
{2\alpha+2\pi m}{\tau_{w}}\right)  \right) \nonumber\\
&  =\frac{2v^{2}\sin^{2}\alpha}{\omega^{2}\tau_{w}^{2}}\sum_{m=-\infty
}^{\infty}\left(  \delta\left(  \omega-\omega_{m}^{+}(\alpha)\right)
+\delta\left(  \omega-\omega_{m}^{-}(\alpha)\right)  \right)  ,
\end{align}
\bigskip where $\omega_{m}^{\pm}(\alpha)=\frac{2\left(  \pm\alpha+\pi
m\right)  }{\tau_{w}}.$

\section{Appendix B}

Using the formula \cite{series}%

\begin{equation}
\cot\theta=\frac{1}{\theta}+\frac{\theta}{\pi}\left(  \sum\limits_{n=-\infty
}^{\infty}\right)  ^{\prime}\frac{1}{n\left(  \theta-n\pi\right)  }%
=\sum\limits_{n=-\infty}^{\infty}\frac{1}{\theta-n\pi}, \label{b1}%
\end{equation}
$\left(  \sum^{\prime}\text{ means }n=0,\text{ excluded}\right)  $one gets:%

\begin{equation}
\frac{\cot\theta-\cot\varphi}{\varphi-\theta}=\sum\limits_{n=-\infty}%
^{+\infty}\frac{1}{(\pi n-\varphi)(\pi n-\theta)}. \label{b2}%
\end{equation}
After differentiation with respect to $\varphi$ we obtain:%

\begin{equation}
\sum\limits_{n=-\infty}^{+\infty}\frac{1}{(\pi n-\varphi)^{2}(\pi n-\theta
)}=\frac{1}{(\varphi-\theta)\sin^{2}\varphi}-\frac{\cot\theta-\cot\varphi
}{(\varphi-\theta)^{2}}. \label{b3}%
\end{equation}

Bromwich, T.J.I. . \emph{An Introduction to the theory of infinite series,
}Second edition,Macmillan, 1942, p.218, (A$_{3}$, A$_{4}$)

Many formulae of this type are derived in: Chrystal, G., \emph{Algebra, an
elementary text book for higher classes of secondary schools and for colleges.
Part II}, Second Edition, 1900, chapter XXX.

\bigskip

\section{Figure Captions}

Figure 1. Note the curves are for a single fixed velocity. The velocity
dependence is contained in the normalization of the frequency scale,
$\omega_{r}=v/R$.

Figure 2. Trajectory of a particle in a cylindrical cell.

Figure 3. Normalized frequency shift for a constant velocity as a function of
normalized applied frequency, $\omega^{\prime}=\omega_{o}R/v,$ for different
values of the damping parameter $r_{o}=R/\lambda.$ Solid curves - results of
the analytic function, equations (\ref{004}) and (\ref{005}). Dotted lines
numerical simulations from ref. \cite{LG}. red - r$_{o}=.2$, green -
r$_{o}=.5$, cyan - r$_{o}=1$, black r$_{o}=2$, blue - r$_{o}=4$, magenta -
r$_{o}=10$.

Figure 4. Normalized velocity averaged frequency shift vs. reduced frequency
$\omega^{\ast}$=$\omega_{o}R/\beta\left(  T\right)  $ for various temperatures
using the temperature-dependent mean free path for $He^{3}$ in $He^{4}$.

Figure 5. Normalized velocity-averaged frequency shift, $\Psi\left(
\omega^{\ast},T\right)  $ vs. temperature, $T$, for various reduced
frequencies $\omega^{\ast}$=$\omega_{o}R/\beta\left(  T\right)  $ using the
temperature-dependent mean free path for $He^{3}$ in $He^{4}$

\begin{thebibliography}{9}                                                                                                %

\bibitem {weinb}Weinberg, S. Proc. XXVI Inter. Conf. on High Energy Physics
(Dallas, Texas) summary talk

\bibitem {jmphinds}Pendlebury, JM and Hinds EA, NIM in Physics Research
\textbf{A440}, 471 (2000)

\bibitem {glrpp}Golub, R and Lamoreaux, SK, Physics Reports, \textbf{237}, 1 (1994),

Harris, PG et al , PRL \textbf{82}, 904 (1999),

LANL EDM collaboration: ''A new search for the neutron electric dipole
moment'', Los Alamos Report LA-UR 02-2331 (2002).

\bibitem {gas}Behr, JA et al, Eur. Phys. J. \textbf{A25}, 685 (2005)

\bibitem {radium}Berg, GPA et al, Nuclear Physics \textbf{A721}, 1107c (2003)

Schulte, E et al, 35th Meeting of the Division of Atomic, Molecular and
Optical Physics, May 25-29, 2004, Tuscon AZ

Ahmad I et al, http://www-mep.phy.anl.gov/atta/research/radiumedm.html

Guest JR et al, Abstract KB.00011 2nd Joint Meeting of the Nuclear Physics
Division of the APS and the Physical Society of Japan, Maui, Hawaii, Sept
18-22 (2005)

\bibitem {radon}Nuss-Warren et al, NIM in Physics Research \textbf{A533}, 275 (2004),

Abstrct KB.00010, 2nd Joint Meeting of the Nuclear Physics Division of the APS
and the Physical Society of Japan, Maui, Hawaii, Sept 18-22 (2005)

\bibitem {xenon}Rosenberry MA and Chupp TE, PRL \textbf{86}, 0031-9007 (2000)

Yoshimi A et al, Physics Letters, \textbf{A304}, 13 (2002) and

Abstract JB.00009, 2nd Joint Meeting of the Nuclear Physics Division of the
APS and the Physical Society of Japan, Maui, Hawaii, Sept 18-22 (2005)

\bibitem {hg}Jacobs JP et al, Phys Rev \textbf{A52}, 3521 (1995)

\bibitem {cumm}Commins, ED; Am. J. Phys. \textbf{59}, 1077 (1991)

\bibitem {jmp}Pendlebury, JM \emph{et al;} Phys. Rev. \textbf{A70}, 032102 (2004)

\bibitem {LG}Lamoreaux, SK and Golub, R; Phys. Rev \textbf{A71}, 032104 (2005)

\bibitem {lanledm}LANL EDM collaboration: ''A new search for the neutron
electric dipole moment'', Los Alamos Report LA-UR 02-2331 (2002).

\bibitem {lametal}Lamoreaux, SK \emph{et al}; Europhys Lett. \textbf{58}, 718 (2002)

\bibitem {mcgreg}McGregor, DD, Phys Rev \textbf{A41}, 2631 (1990)

\bibitem {papoul}Papoulis, A, \emph{Probability, Random Variables and
Stochastic Processes, }McGraw Hill, NY \ (1965)

\bibitem {harrjmp}Harris, PG and Pendlebury, JM, \emph{ }arxiv.org:
physics/0510134, ct. 2005, Phys Rev \textbf{A73}, 014101 (2006)

\bibitem {baymebn}Baym, G and Ebner, C; Phys. Rev. \textbf{164}, 235 (1967)

\bibitem {series}Gradshteyn, I.S. and Ryzhik, I.M., \emph{Table of Integrals,
Series and Products, }Academic Press, 1965, 1.421.3, p. 36.
\end{thebibliography}
\end{document}